\documentstyle[12pt,psfig]{article} 
\newcommand{\T}[1]{{\mathrm{#1}}}

\newcommand{\gsim}{\buildrel > \over {_\sim}}
\newcommand{\lsim}{\buildrel < \over {_\sim}}

\newcommand{\Slash}[1]{{#1}\!\!\!/}

\newcommand{\etal}{{\em et al.}}

\newcommand{\PRD}[3]{Phys.\ Rev.\ D {\bf {#1}}, {#2} ({#3})}
\newcommand{\PRL}[3]{Phys.\ Rev.\ Lett.\ {\bf {#1}}, {#2} ({#3})}
\newcommand{\NPA}[3]{Nucl.\ Phys.\ A {\bf {#1}}, {#2} ({#3})}
\newcommand{\NPB}[3]{Nucl.\ Phys.\ B {\bf {#1}}, {#2} ({#3})}
\newcommand{\PLB}[3]{Phys.\ Lett.\ B {\bf {#1}}, {#2} ({#3})}
\newcommand{\Rep}[3]{Phys.\ Rep.\ {\bf {#1}}, {#2} ({#3})}
\newcommand{\ZPC}[3]{Z. Phys.\ C {\bf {#1}}, {#2} ({#3})}
\newcommand{\ZPA}[3]{Z. Phys.\ A {\bf {#1}}, {#2} ({#3})}

\newcommand{\ppp}[1]{%
        \setbox0=\hbox{#1}%
        \kern-.02em\copy0\kern-\wd0
        \kern+.04em\copy0\kern-\wd0
        \kern-.02em\raise.0217em\box0}

\begin{document}  

\renewcommand{\thefootnote}{\fnsymbol{footnote}}

\begin{flushright}
       TUM-T39-98-20 \\
       JYFL-13/98 \\
       hep-ph/9808330 \\
       \today
\end{flushright}

\vspace*{2cm}

\begin{center}
{\Large\bf \setcounter{footnote}{1}
Nuclear Quark and Gluon Distributions \\
in Coordinate Space\footnote{Work supported in part by BMBF}}
\end{center}

\bigskip
\bigskip

\setcounter{footnote}{0}
\centerline{M. V\"anttinen\footnote{Alexander von Humboldt fellow},
G. Piller, L. Mankiewicz and W. Weise}

\begin{center}
{\em Physik Department, Technische Universit\"{a}t M\"{u}nchen, \\
D-85747 Garching, Germany}
\end{center}

\medskip
{\centerline {K. J. Eskola}}

\begin{center}
{\em Department of Physics, University of Jyv\"askyl\"a, \\ 
P.O.Box 35, FIN-40351 Jyv\"askyl\"a, Finland}
\end{center}

\renewcommand{\thefootnote}{\arabic{footnote}}
\setcounter{footnote}{0}

\vspace*{3cm}
\begin{abstract} 
\bigskip

\noindent
In coordinate space, quark and gluon distributions of the nucleon
are defined as correlation functions involving two field operators
separated by a light-cone distance $y^+ = 2l$. We study the nuclear
modifications of these distributions. The largest effect
is a strong depletion of parton distributions (shadowing) at large
longitudinal distances, which starts
for all parton species at $l=2$ fm, i.e.\ at the average
nucleon-nucleon separation in nuclei. On the other hand, the nuclear radius
does not play a significant role.
At $l \lsim 1$ fm, nuclear modifications of parton distributions
are small. The intrinsic structure of individual nucleons is
evidently not very much affected by nuclear binding. In particular,
there is no evidence for a significant increase of the quark or
gluon correlation length in bound nucleons.
\medskip


\end{abstract}
\thispagestyle{empty}
\newpage

\section{Introduction}

A significant difference between nucleon and nuclear structure
functions was first observed in deep inelastic scattering by the
EMC collaboration \cite{EMC-observation}. Since then, considerable
experimental and theoretical efforts have been devoted to detailed
investigations of nuclear modifications of parton distributions
(for a review see e.g.\ \cite{Arneodo}). Their phenomenological
discussion has been carried out almost entirely in momentum space. 
However, interesting insights can be obtained also in coordinate
space. Here the parton distributions are defined in leading twist
accuracy as correlation functions involving two quark or gluon
field operators, separated by a light-cone distance $y^+$
\cite{Collins:1982uw,BalitskyBraun}.
In deep inelastic scattering from nuclei as viewed in the 
laboratory frame, where the target is at rest, the longitudinal
distance $y^+/2$ involved in the parton correlation functions can
be compared with typical length scales provided by the nucleus.
This offers new possibilities for extracting information on the 
nature of nuclear effects in parton distributions
\cite{Llewellyn,Hoyer-MV}.

To demonstrate the relevance of coordinate-space distances
in nuclear parton distributions we focus on
deep inelastic scattering. Consider first the scattering from a
free nucleon with momentum $P$ and invariant mass $M$ in the
laboratory frame. The momentum transfer $q$, carried by the exchanged 
virtual photon, is taken in the (longitudinal) $\hat 3$-direction, 
$q^{\mu}=(\nu,\vec 0_{\perp},\sqrt{\nu^2 + Q^2})$, with $Q^2 = -q^2$. 
In the Bjorken limit, $\nu^2 \gg Q^2 \gg M^2$ with $x=Q^2/(2 M \nu)$
fixed, the light-cone components of the photon momentum,
$q^{\pm}=q^0 \pm q^3$, are $q^+ \simeq 2 \nu$ and $q^- \simeq - Mx$.
All information about the response of the target to the high-energy
interaction is in the hadronic tensor
\begin{equation} \label{eq:WJJ_ST}
  W_{\mu\nu} (q,P) \sim 
  \int d^4 y \,e^{i q\cdot y} \, 
  \langle P| J_{\mu}(y) J_{\nu}(0)|P\rangle,
\end{equation}
defined as the Fourier transform of a product of the electromagnetic
currents $J_{\mu}$, with its expectation value taken between the
nucleon states. Using 
\begin{eqnarray}
  q\cdot y = \frac{1}{2}\left(q^+ y^- + q^- y^+\right) - 
             \vec q_{\perp} \cdot \vec y_{\perp}
           = \nu \,y^- - \frac{Mx}{2} \,y^+,
\end{eqnarray}
one obtains the following coordinate-space resolutions along the 
coordinates $y^{\pm} = t \pm y^3$:
\begin{equation} \label{eq:typical_y}
  \delta y^- \sim \frac{1}{\nu}
  \quad\mbox{and} \quad 
  \delta y^+ \sim  \frac{1}{Mx}.
\end{equation}
At $y^-=0$ the current correlation function in Eq.\ 
(\ref{eq:WJJ_ST}) is not analytic since it vanishes for
$y^+ y^- - {(\vec y_{\perp})}^2 < 0$ because of causality
(see e.g.\ \cite{Muta:1987mz}). Indeed in perturbation theory it
turns out to be singular at $y^-=0$. Assuming that the integrand
in (\ref{eq:WJJ_ST}) is an analytic function of $y^-$ elsewhere, 
this implies that $W_{\mu\nu}$ is dominated for
$q^+ \rightarrow \infty$  by contributions from $y^-= 0$. Using
causality one then finds that, in the transverse plane, only 
contributions from ${(\vec y_{\perp})}^2 \simeq 1/Q^2$ are relevant: 
in deep inelastic scattering the hadronic tensor is dominated 
by contributions from the light cone, i.e.\ $y^2 = 0$.

Furthermore, Eq.\ (\ref{eq:typical_y}) suggests that along the light
cone, one is dominantly probing larger distances as $x$ is decreased.
It has been shown that such a behavior is consistent with approximate 
Bjorken scaling \cite{scaling}. An analysis of nucleon structure
functions in coordinate space as carried out in Section
\ref{section:free-nucleon} confirms the above conjecture. In the
Bjorken limit the dominant contributions to the hadronic tensor at
small $x$ come from light-like separations of order $y^+ \sim 1/(Mx)$
between the electromagnetic currents.

In the laboratory frame our considerations imply that deep inelastic
scattering involves a longitudinal correlation length  
\begin{equation} \label{eq:l_z}
  y^3 \simeq \frac{y^+}{2} \equiv l 
\end{equation}
of the virtual photon. Consequently, large longitudinal distances 
are important in the scattering process at small $x$. This can also
be deduced in the framework of time-ordered perturbation theory
(see e.g. \cite{TOPTh}), where the typical propagation length of
hadronic configurations present in the interacting photon is
$y^3 \sim 1/(2Mx)$, in accordance with our discussion.

\begin{figure}
\centerline{\psfig{figure=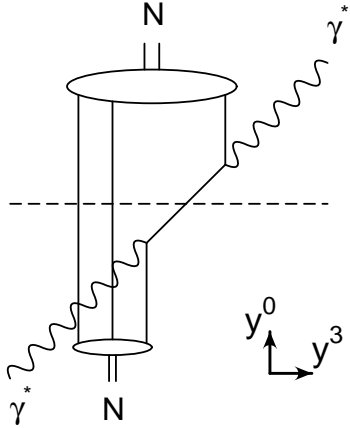,width=5cm}}
\caption{A Feynman diagram from Ref.~\protect\cite{Hoyer-MV} 
illustrating the space-time pattern of deep inelastic scattering.} 
\label{diagram}
\end{figure}

The space-time pattern of deep inelastic scattering is illustrated
in Fig.~\ref{diagram} in terms of the imaginary part of the forward
Compton amplitude: the virtual photon interacts with a quark or
antiquark which is displaced a distance $y^+$ along the light cone. 
The characteristic laboratory frame correlation length $l$ is one 
half of that distance.
This behavior is naturally implemented in coordinate-space  
(or so-called  Ioffe-time) distribution functions. They are 
related to momentum-space distributions through Fourier
transformation and thus select contributions to the scattering
process which result from specific light-cone distances $y^+$. 
For example, the charge-conjugation-even quark distribution 
in coordinate space reads:
\begin{eqnarray} 
  {\cal Q}(y^+,Q^2) &\equiv& 
  \int_0^1 {d}x\,  
  \left [q(x,Q^2) + \bar q(x,Q^2) \right] \,
  \sin \left( \frac{M y^+}{2}\,x \right) \, . 
\end{eqnarray}
Here $q(x,Q^2)$ and $\bar q(x,Q^2)$ are the quark and antiquark
momentum-space distribution functions which depend on Bjorken $x$
and on the momentum scale $Q^2$. At lowest order in the strong
coupling constant $\alpha_s$, the distribution ${\cal Q}$ is
related to the structure function $F_2$, which is measured in
deep inelastic scattering, through:\footnote{In higher orders
in $\alpha_s$ this is a matter of scheme conventions.}
\begin{eqnarray} \label{eq:Ioffe_F2}
  {\cal F}_2(y^+,Q^2) = \sum_f e_f^2 \,{\cal Q}_f(y^+,Q^2) =
  \int_0^1 \frac{ d x}{x} \,F_{2} (x,Q^2) 
  \sin \left( \frac{My^+}{2} \,x \right) \, ,  
\end{eqnarray}
where $e_f$ is the fractional electromagnetic charge of a quark 
with flavor $f$. 

If one compares the longitudinal correlation length $l$ from
Eq.\ (\ref{eq:l_z}) with the average nucleon-nucleon distance
in the nucleus, $d \approx 2$ fm, one can distinguish two
different kinematic regions:
\begin{enumerate}
\item[(i)] 
At small distances, $l < d$, the virtual photon scatters incoherently
from the individual hadronic constituents of the nuclear target. 
Possible  modifications of ${\cal Q}(y^+)$ in this region are 
caused by bulk nuclear effects such as binding and Fermi motion.

\item[(ii)] 
At larger distances, $l > d$, it is likely that several nucleons 
participate collectively in the interaction. Modifications of
${\cal Q}(y^+)$ are now expected to come from the coherent
scattering of the photon on several nucleons in the target. Using
$l \sim 1/(2 M x)$, this region corresponds to the kinematic domain
$x \lsim 0.05$. 
\end{enumerate}
This suggests that the nuclear modifications seen in coordinate-space
distributions will be  quite different in the regions $l > 2$ fm and
$l < 2$ fm.

In the following we first recall, in Section \ref{section:cspace},
the definition of coordinate-space distribution functions and review
their relation to QCD operators. In Section \ref{section:free-nucleon}
we discuss the coordinate-space distribution functions of free nucleons. 
Nuclear modifications of quark and gluon distributions are investigated
in Section \ref{section:nuclear-effects}. A short summary is given in
Section \ref{section:summary}.

\section{Coordinate-space distribution functions \label{section:cspace}}

It is useful to express coordinate-space distributions in terms of
a dimensionless variable. For this purpose let us introduce the
light-like vector $n^{\mu}$ with $n^2 = 0$ and $P\cdot n = P^0 - P^3$.
As discussed in the introduction, dominant contributions to Eq.\ 
(\ref{eq:WJJ_ST}) come from the vicinity of the light cone, where
$y$ is approximately parallel to $n$. The dimensionless variable
$z = y \cdot P$ then plays the role of a coordinate conjugate to
Bjorken $x$. It is useful to bear in mind that the value $z = 5$
corresponds in the laboratory frame to a light-cone distance 
$y^+ =  2 z/M \approx 2$ fm or, equivalently, to a longitudinal
distance $l \equiv y^+/2 \approx 1$ fm.

In accordance with the charge conjugation ($C$) properties of
momentum-space quark and gluon distributions, one defines
coordinate-space distributions by \cite{evolution}
\begin{eqnarray}
  \label{eq:Coordinate_1}
  {\cal Q}(z,Q^2) &\equiv& \int_0^1 {d}x
  \, \left [q(x,Q^2) + \bar q(x,Q^2) \right]\,\sin (z \,x),   
  \\
  \label{eq:Coordinate_2}
  {\cal Q}_{v}(z,Q^2) &\equiv& \int_0^1 {d}x\, 
  \left[q(x,Q^2) - \bar q(x,Q^2) \right]\, \cos (z \,x),
  \\ 
  \label{eq:Coordinate_3}
  {\cal G}(z,Q^2) &\equiv& \int_0^1 {d}x\, 
  x\,g(x,Q^2)\, \cos (z \,x), 
\end{eqnarray}
where $q$, $\bar q$ and $g$ are the momentum-space quark, antiquark 
and gluon distributions, respectively. Flavor degrees of freedom
are suppressed here.

In leading twist accuracy, the coordinate-space distributions 
(\ref{eq:Coordinate_1}--\ref{eq:Coordinate_3}) 
are related to forward matrix elements of non-local QCD operators
on the light cone \cite{Collins:1982uw,BalitskyBraun}:
\begin{eqnarray} 
  \label{eq:Coord_Op_Q}
  {\cal Q}(z,Q^2) 
  &=& \frac{1}{4 i P\cdot n} \,
  \langle P| \overline \psi(y) \,
  \Slash{n} 
  \Gamma(y) \,\psi(0)|P\rangle_{Q^2}
  - (y \leftrightarrow -y),
  \\
  \label{eq:Coord_Op_Qv}
  {\cal Q}_{v}(z,Q^2) 
  &=& \frac{1}{4 P\cdot n} \,
  \langle P| \overline {\psi}(y) \,
  \Slash{n} 
  \Gamma(y) \,\psi(0)
  |P\rangle_{Q^2}
  + (y \leftrightarrow -y), 
  \\
  \label{eq:Coord_Op_G}
  {\cal G}(z,Q^2) 
  &=& 
  n^{\mu} n^{\nu} 
  \frac{1}{2 (P\cdot n)^2}
  \langle P| G_{\mu\lambda}(y) \,\Gamma(y)\,
  G^{\lambda}_{\,\,\nu}(0)|P\rangle_{Q^2}.
\end{eqnarray}
Here $\psi$ denotes the quark field and $G_{\mu\nu}$ the gluon
field strength tensor. The path-ordered exponential
\begin{equation}
  \Gamma(y) = {\T P}\exp
  \left[ ig\,y^{\mu} \int_0^1 {d} \lambda  \,A_{\mu}(\lambda y)  
  \right] \, ,
\end{equation}
where $g$ denotes the strong coupling constant and $A^{\mu}$ the
gluon field, ensures gauge invariance of the parton distributions.
Note that an expansion of the right-hand side of Eqs.\
(\ref{eq:Coordinate_1}--\ref{eq:Coordinate_3}) and
(\ref{eq:Coord_Op_Q}--\ref{eq:Coord_Op_G}) around $y = 0$ leads
to the conventional operator product expansion for parton
distributions \cite{Muta:1987mz}.

The functions ${\cal Q}(z)$, ${\cal Q}_v(z)$ and ${\cal G}(z)$ 
characterize the mobility of partons in coordinate space. Consider, for
example, the quark distributions ${\cal Q}(z)$ and ${\cal Q}_{v}(z)$.
The matrix elements in (\ref{eq:Coord_Op_Q},\ref{eq:Coord_Op_Qv}) have
an obvious physical interpretation:  they measure the overlap between
the nucleon ground state and a state in which one quark has been
displaced along the light cone from $0$ to $y$. Antiquarks undergo this
sequence in opposite order.

\section{Parton distribution functions
of free nucleons \label{section:free-nucleon}}

\begin{figure}
\input{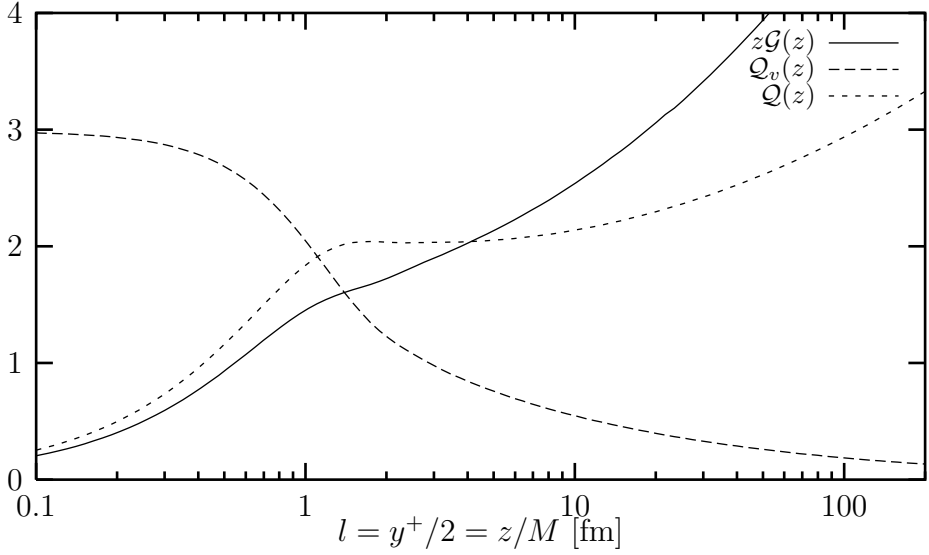}
\caption{Coordinate-space quark and gluon distributions resulting
from the CTEQ4L parametrization of momentum-space distributions,
taken at a momentum transfer $Q^2 = 4$ GeV$^2$. A sum over the $u$
and $d$ quarks is implied in the functions ${\cal Q}_v$ and ${\cal Q}$.}
\label{fig:freenucleon}
\end{figure}

In this section we discuss the properties of coordinate-space
distribution functions of free nucleons. Examples of the distributions 
(\ref{eq:Coord_Op_Q}--\ref{eq:Coord_Op_G}) using
the CTEQ4L parametrization \cite{CTEQ4L} of momentum-space quark
and gluon distributions taken at a momentum scale $Q^2 = 4$ GeV$^2$,
are shown in Fig.~\ref{fig:freenucleon}.

\begin{figure}
\input{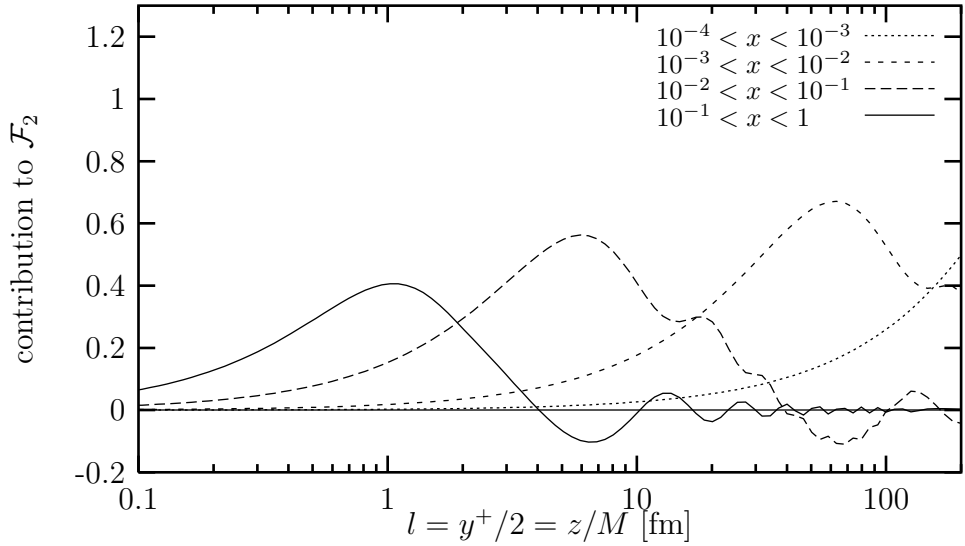}
\caption{Contributions from different regions in $x$ to the ${\cal F}_2$
combination of coordinate-space quark and antiquark distributions. The
CTEQ4L parametrization of momentum-space distributions, taken at a
momentum scale $Q^2 = 4$ GeV$^2$, has been used.}
\label{fig:xregions}
\end{figure}

Some general features can be observed: the $C$-even quark distribution
${\cal Q}(z)$ rises at small
values of $z$, develops a plateau at $z \gsim 5$, and then exhibits a
slow rise at very large $z$.
At $z \lsim 5$, the function $z \,{\cal G}(z)$ behaves similarly as
${\cal Q}(z)$. For $z \gsim 5$, $z \,{\cal G}(z)$ rises somewhat
faster than ${\cal Q}(z)$. The $C$-odd (or valence) quark distribution
${\cal Q}_v(z)$ starts with  a finite value at small $z$, then begins to
fall at $z \simeq 3$, and vanishes at large $z$. Recall that, in the
laboratory frame, the scale $z \simeq 5$ at which a significant change
in the behavior of coordinate-space distributions occurs, represents 
a longitudinal distance comparable to the typical size of a nucleon. 

At $z < 5$ the coordinate-space distributions are determined by average
properties of the corresponding momentum-space distribution functions
as expressed by their first few moments \cite{firstfew}. For example,
the derivative of the $C$-even quark distribution ${\cal Q}(z)$ taken
at $z = 0$ equals the fraction of the nucleon light-cone momentum
carried by quarks. The same is true for the gluon distribution
$z \,{\cal G}(z)$ (the momentum fractions carried by quarks and by
gluons are in fact approximately equal, a well-known experimental fact).  
At $z > 10$ coordinate-space distributions are determined by the small
$x$ behavior of the corresponding momentum-space distributions. 
Assuming, for example, $q(x) \sim x^{\alpha}$  for $x < 0.05$ implies
${\cal Q}(z) \sim z^{-\alpha - 1}$ at $z > 10$. Similarly, the small
$x$ behavior $g(x) \sim x^{\alpha}$ leads to  
$z \,{\cal G}(z) \sim z^{ - \alpha - 1}$ at large $z$. For typical
values  of $\alpha$ as suggested by Regge phenomenology \cite{Regge}
one obtains ${\cal Q}_{v} \sim z^{-0.5}$ while ${\cal Q}(z)$ and
$z \,{\cal G}(z)$ become constant at large $z$. 

The following argument makes it plausible that $z \,{\cal G}(z)$
behaves in a similar way as ${\cal Q}(z)$ at large $z$: In leading
order in the strong coupling constant $\alpha_s$, gluons enter
through $Q^2$ evolution in the flavor singlet channel. For the
flavor-singlet quark distribution at momentum scale $Q^2$ one
has\footnote{The complete leading order DGLAP evolution equations in  
coordinate space can be found for example in \cite{evolution}.} 
\begin{eqnarray}
  \label{eq:evo}
  \frac{\partial{\cal Q}(z,Q^2)}{\partial \ln Q^2} 
  &=&  
  - \frac{\alpha_s}{2 \pi}  
  \left\{
  C_F \int_0^1 du \,K_{\T{\cal QQ}}(u) \, {\cal Q}(u z,Q^2) 
  \right.
  \nonumber
  \\
  &&
  \hspace{4cm}
  \left.
  + 
  N_f \int_0^1 du \,K_{\T{\cal QG}}(u) \, z \,{\cal G}(u z,Q^2) 
  \right\},
\end{eqnarray}
where $C_F = 4/3$ and $N_f$ is the number of active flavors. 
The quark-quark and quark-gluon splitting functions read:
\begin{eqnarray}
  K_{\T{\cal QQ}}(u) &=& \frac{1}{2} \delta(1-u) - (1-u) - 
  2 \frac{u}{1-u} - \delta(1-u) \int_0^1 du' \frac{1-u'}{u'}, \\
  K_{\T{\cal QG}}(u) &=& -\frac{1}{3} (1-u)
  \, \left[ 2(1-u)^2 + 3\,u \right].
\end{eqnarray} 
We are primarily interested in the region of small Bjorken $x$, 
$x < 0.01$, which corresponds to large $z \simeq 1/(2 x)$. Looking
at Fig.~\ref{fig:freenucleon} we see that the integrals in Eq.\ 
(\ref{eq:evo}) receive their dominant contributions from the region
$u z > 5$ or  $u > u_0 = 10 \,x$. In the relevant interval
$u_0 < u \leq 1$ the quark and gluon distributions ${\cal Q}(uz)$
and $uz \,{\cal G}(uz)$ are smooth functions which can be replaced
approximately by their values at a point $u=\bar u$ within the
interval. Neglecting corrections from $u < u_0 $ gives: 
\begin{eqnarray} \label{eq:approx}
  \frac{\partial{\cal Q}(z,Q^2)}{\partial \ln Q^2} 
  \approx
  - \frac{\alpha_s}{2 \pi} 
  \left\{
  C_F \,
  {\cal Q}(\bar u z,Q^2) \int_{u_0}^1 du \,K_{\T{\cal QQ}}(u)  
  + 
  N_f 
  \, \bar u z \,{\cal G}(\bar u z,Q^2) \int_{u_0}^1 \frac{du}{u} \,
  K_{\T{\cal QG}}(u) 
  \right\}.
\end{eqnarray}
We find indeed that at small values of $x$ or, equivalently, at
large $z$ the gluon distribution enters in terms of $z \,{\cal G}(z)$.
For $x < 0.05$, which corresponds to $u_0 < 0.4$, the integral over
the gluon splitting function dominates increasingly over the quark
contribution. This reflects the well-established dominance of gluons
in the QCD evolution at small $x$ \cite{dominance}. Note that the
strong rise of quark and gluon distribution functions at small $x$
and large $Q^2$ as observed at HERA leads to an increase of the
corresponding coordinate-space distributions at large $z$. In this
kinematic region, corrections involving derivatives of the gluon
distribution have to be included in Eq.\ (\ref{eq:approx}).
  
Finally we illustrate the relevance of large distances in deep inelastic
scattering at small $x$ as discussed in the introduction. In
Fig.~\ref{fig:xregions} we show contributions to the structure function
$F_2$ in coordinate space which result from different regions of 
Bjorken $x$. We confirm indeed that contributions from large 
distances $\sim 1/(Mx)$ dominate at small $x$.

\section{Nuclear parton distribution
functions \label{section:nuclear-effects}}

A detailed analysis of nuclear parton distribution functions
in momentum space was performed recently in \cite{Eskola}
(for earlier investigations see also \cite{Frankfurt,Eskola-NPB400}).
In this work a set of nuclear parton distributions at an initial momentum
scale $Q_0^2 = 2.25$ GeV$^2$ was determined by using data from  
deep inelastic lepton-nucleus scattering and Drell-Yan dilepton
production in proton-nucleus collisions. Important constraints
were imposed by baryon number and momentum conservation. 
Based on the complete set of these data, a lowest order 
DGLAP evolution analysis was performed in order to extract
separately the quark and gluon content of the nuclear
distributions. Good agreement with present experimental data
for the ratio of the nuclear to the free nucleon structure
function, $F_2^{\T A} / F_2^{\T N}$, including its dependence
on the momentum transfer $Q^2$, was achieved. 

\begin{figure}
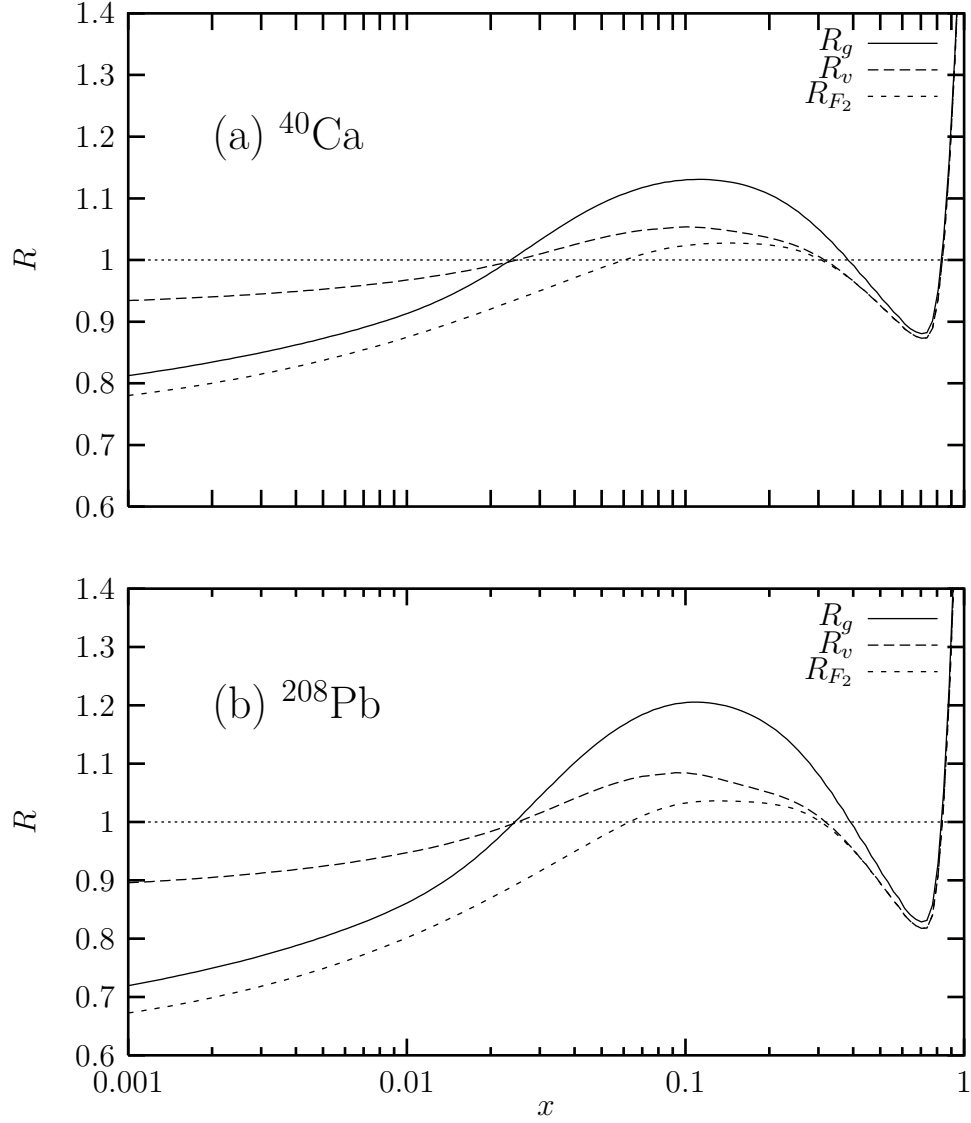

\input{mSAEF.Ca.tex}
\input{mSAEF.Pb.tex}
\caption{Momentum-space ratios at $Q^2 = 4$ GeV$^2$ for gluon
distributions, valence-quark distributions, and the $F_2$ structure
function in (a) $^{40}$Ca and (b) $^{208}$Pb according to
Ref.~\protect\cite{Eskola}.}
\label{fig:momspace-Ca-Pb}
\end{figure}

We shall consider the following ratios of quark and gluon 
distribution functions normalized to the number of nucleons 
in the target:  
\begin{eqnarray}
  R_{F_2} (x,Q^2) &=& \frac{F_2^{\T A}(x,Q^2)}{F_2^{\T N}(x,Q^2)} 
  = \frac{\sum_f e_f^2 \left[ q^{\T A}_f(x,Q^2) + 
  \bar q^{\T A}_f(x,Q^2)\right]}
  {\sum_f e_f^2 \left[ q^{\T N}_f(x,Q^2) + \bar q^{\T N}_f(x,Q^2)\right]},
  \\
  R_v(x,Q^2) &=& \frac{\sum_f \left[ 
  q^{\T A}_f(x,Q^2) - \bar q^{\T A}_f(x,Q^2) \right]}
  {\sum_f \left[ q^N_f(x,Q^2)- \bar q^N_f(x,Q^2) \right]},
  \\
  R_g(x,Q^2) &=& \frac{ g^{\T A}  (x,Q^2)}{ g^{\T N} (x,Q^2)}. 
\end{eqnarray}
In Fig.~\ref{fig:momspace-Ca-Pb} typical results from
Ref.~\cite{Eskola} are shown for $^{40}$Ca and $^{208}$Pb
taken at $Q^2 = 4$ GeV$^2$. The ratios shown here and below have been
obtained using the GRV-LO parton distributions \cite{GRV-LO}.
It was recently shown in Ref.\ \cite{Eskola-new} that the
results are insensitive to the choice of the parton distribution
set.  

The behavior of the structure function ratio $R_{F_2}$ and its
interpretations are familiar from earlier experimental and
theoretical work (see e.g.\ \cite{Arneodo}). At $x<0.1$ the ratio
is smaller than one, i.e.\ there is nuclear shadowing,
$F_2^{\T A} < F_2^{\T N}$. A small enhancement ("antishadowing")
of $F_2^{\T A}$ is found at $x\sim 0.1$. In the region
$0.2 < x < 0.7$ a significant depletion of $R_{F_2}$ (the "EMC effect")
can be seen. The strong rise of $R_{F_2}$ at
$x>0.8$ is caused by Fermi motion. Qualitatively similar effects
were obtained for the first time in Ref.\ \cite{Frankfurt}
for gluon and valence quark distributions.

In Ref.\ \cite{Eskola} it was assumed that
at the initial momentum scale $Q_0^2$ shadowing is of similar size
for gluons as for the structure function $F_2$, i.e.\
$R_g (x,Q_0^2) \approx R_{F_2}(x,Q_0^2)$ for $x<0.01$.\footnote{Further
assumptions, which are not of immediate relevance for our discussion
here, are explained in the original paper \cite{Eskola}.} Because of
momentum conservation, gluon shadowing at small $x$ implies antishadowing
of nuclear gluon distributions at larger $x$. An analysis of the $Q^2$
dependence of the structure function ratio $F_2^{\T{Sn}}/F_2^{\T{C}}$
\cite{Gousset} has shown that gluon antishadowing in Sn is located
in the region $0.03 < x < 0.4$. In \cite{Eskola} it was assumed that
this applies also to other nuclei. Antishadowing was then found to reach
its maximum at $x\simeq 0.15$, the maximum being $13\%$ for $^{40}$Ca
and $20\%$ for $^{208}$Pb, as shown in Fig.~\ref{fig:momspace-Ca-Pb}.

The E772 Drell-Yan data show negligible antishadowing of the nuclear
quark sea \cite{DY-data}. To account for the small but significant
antishadowing of $F_2$ at $x\simeq 0.15$, nuclear valence quarks have
to be enhanced in this region. Baryon number conservation then implies
shadowing at $x < 0.1$ also for valence quarks.

Finally let us mention that in Ref.\ \cite{Eskola} the  
gluon and sea quark distributions were assumed to show, 
at $x>0.4$, an ``EMC effect'' similar to $F_2$. Since both 
distributions are small in this domain, this assumption is 
however of minor importance for our study.

\subsection{Coordinate-space results}

\begin{figure}
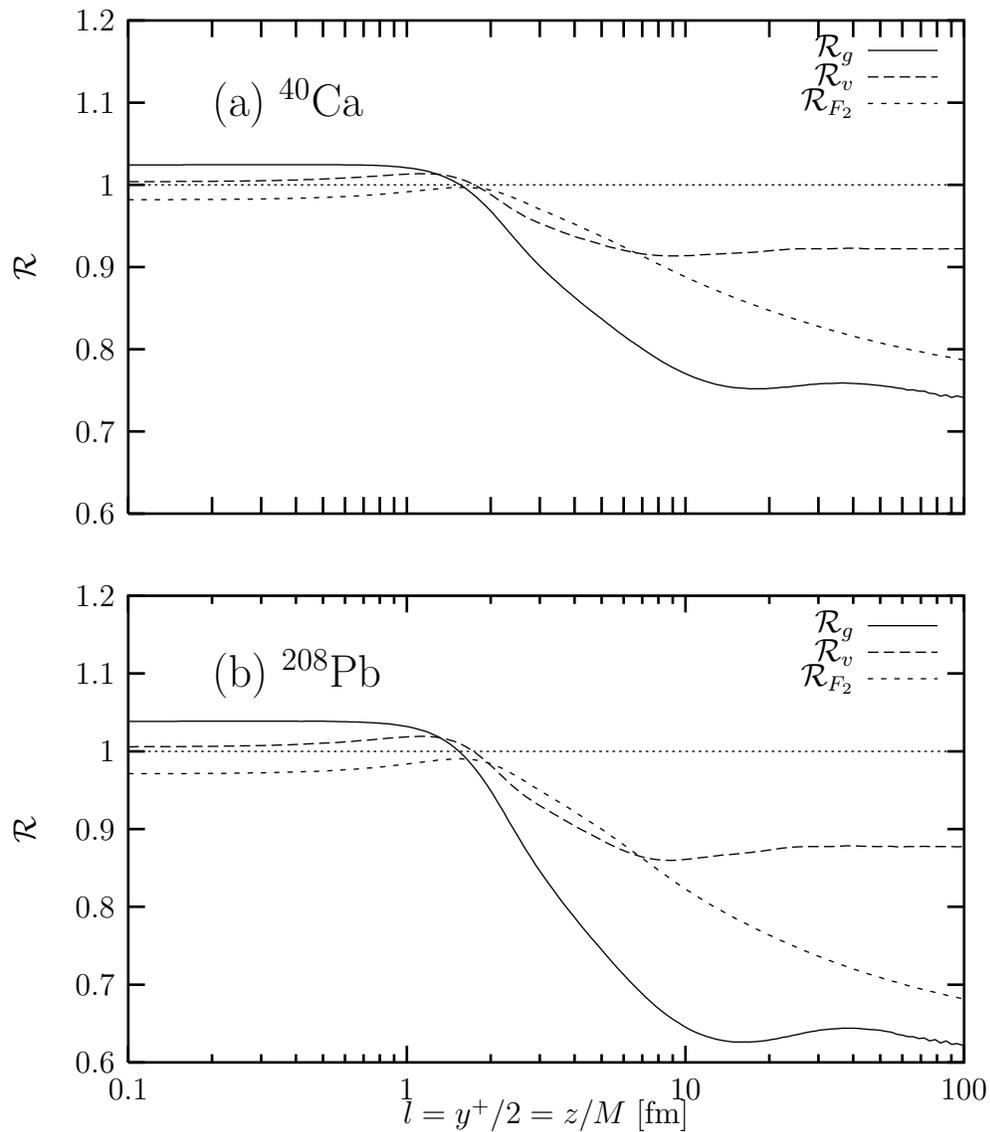

\input{cSAEF.Ca.tex}
\input{cSAEF.Pb.tex}
\caption{Coordinate-space ratios at $Q^2 = 4$ GeV$^2$ for gluon
distributions, valence-quark distributions, and the $F_2$ structure
function in (a) $^{40}$Ca and (b) $^{208}$Pb.}
\label{fig:coordspace-Ca-Pb}
\end{figure}

Using the momentum-space distributions from \cite{Eskola} we have
calculated the corresponding ratios for nuclear and nucleon 
coordinate-space distribution functions:
\begin{eqnarray}
  {\cal R}_{F_2}(z,Q^2) &=& 
  \frac{
  \int_0^A \frac{d x}{x} \,F_{2}^{\T A}(x,Q^2) \sin(z \,x)
  }
  {\int_0^1 \frac{d x}{x}  \,F_{2}^{\T N}(x,Q^2) \sin(z \,x)
  }
  = 
  \frac{\sum _f e_f^2 \,{\cal Q}^{\T A}_f(z,Q^2)}
  {\sum _f e_f^2 \,{\cal Q}^{\T N}_f(z,Q^2)}, 
  \\
  {\cal R}_{v}(z,Q^2) &=& 
  \frac{{\cal Q}^{\T A}_{v}(z,Q^2)}
  {{\cal Q}^{\T N}_{v}(z,Q^2)},
  \\
  {\cal R}_{\cal G}(z,Q^2) &=& 
  \frac{z {\cal G}^{\T A}(z,Q^2)}{z {\cal G}^{\T N}(z,Q^2)}.  
\end{eqnarray}
Results for $^{40}$Ca and $^{208}$Pb are presented in
Fig.~\ref{fig:coordspace-Ca-Pb}.

At large longitudinal distances $l > 2$ fm a strong depletion of
nuclear parton distributions is found. As discussed in Section
\ref{section:free-nucleon}, the asymptotic behavior of
coordinate-space distributions at large $l$ is determined by the
small $x$ asymptotics of the momentum-space distributions. The
coordinate-space ratio at large distances thus corresponds to the
nuclear shadowing seen in momentum space at $x < 0.1$. 
Note, however, that at finite distances the coordinate-space
distributions correspond to a weighted integral of momentum-space
distributions. A given region in momentum space is not simply mapped into
a particular finite region in coordinate space. It would, for
example, be incorrect to associate the ${\cal R}_g<1$ region in 
Fig.~\ref{fig:coordspace-Ca-Pb} exclusively with the $R_g<1$
region in Fig.~\ref{fig:momspace-Ca-Pb}. Instead, the antishadowing
region of momentum space also contributes significantly up to
$l \lsim 10$ fm. We shall illustrate the mapping of momentum-space
regions into coordinate space in more detail in
Section \ref{mapping}.

Shadowing in deep inelastic scattering in the laboratory frame can
be interpreted as being caused by the coherent scattering of hadronic
fluctuations of the exchanged virtual photon with several nucleons
in the target. This picture is nicely consistent with the observation,
in Fig.~\ref{fig:coordspace-Ca-Pb}, that the nuclear depletion in
coordinate space begins at distances close to the average
nucleon-nucleon separation in nuclei, $l \sim 2$ fm. In momentum
space, the Bjorken $x$ value where shadowing begins does not have
such a suggestive direct relation to a physical scale.

Another interesting observation is that in coordinate space,
shadowing sets in at approximately the same value of $l$ for all
sorts of partons. In momentum space, shadowing is found to start at 
different values of $x$ for different distributions.

At $l \lsim 1$ fm, nuclear modifications of parton distributions are 
small. In this region, deep inelastic scattering proceeds from the
individual hadronic constituents in the target nucleus. The intrinsic
structure of individual nucleons is evidently not very much affected
by nuclear binding.

In the limit $l\rightarrow 0$ the ratio ${\cal R}_{F_2}$ compares
the fractions of target momentum carried by quarks in nuclei and in
free nucleons, and ${\cal R}_g$ compares the momentum fractions
carried by gluons. Obviously, ${\cal R}_{F_2}$ and ${\cal R}_g$ are
correlated because of momentum conservation. For valence quarks,
baryon number conservation demands ${\cal R}_{v} = 1$ for
$l\rightarrow 0$.

We find that at small distances $l \leq 1$ fm the gluon distribution
of bound nucleons is enhanced \cite{Frankfurt} by a small amount
(about 3\% for Ca)
with respect to free nucleons. The quark distribution is depleted
accordingly. This result is, however, consistent with zero within
experimental errors (a rigorous error analysis was not performed in
Ref.\ \cite{Eskola}, but the error of the total quark momentum integral
has been analyzed in Ref.\ \cite{Arneodo}).

In early discussions of the EMC effect \cite{early-EMC}, the possibility
of an increased correlation length of partons in nuclei has
been suggested. Within the accuracy of present data, this cannot be
verified. Actually it is difficult to give a precise definition of an
average correlation length, since the
coordinate-space parton distributions are not normalizable and thus
cannot be interpreted as parton number densities.

\subsection{Mapping of momentum-space regions
into coordinate space \label{mapping}}
  
To disentangle the relationship between nuclear effects of parton 
distributions in momentum and coordinate space, it is instructive 
to consider hypothetical momentum-space ratios in which only one
of the observed effects (shadowing, antishadowing or the EMC effect)
is present. The appropriate Fourier transformations will then show
which coordinate-space distances are most closely associated with
these momentum-space effects. In the following we consider $^{40}$Ca,
fixing the momentum scale at $Q^2 = 4$ GeV$^2$ again. 
 
\begin{figure}
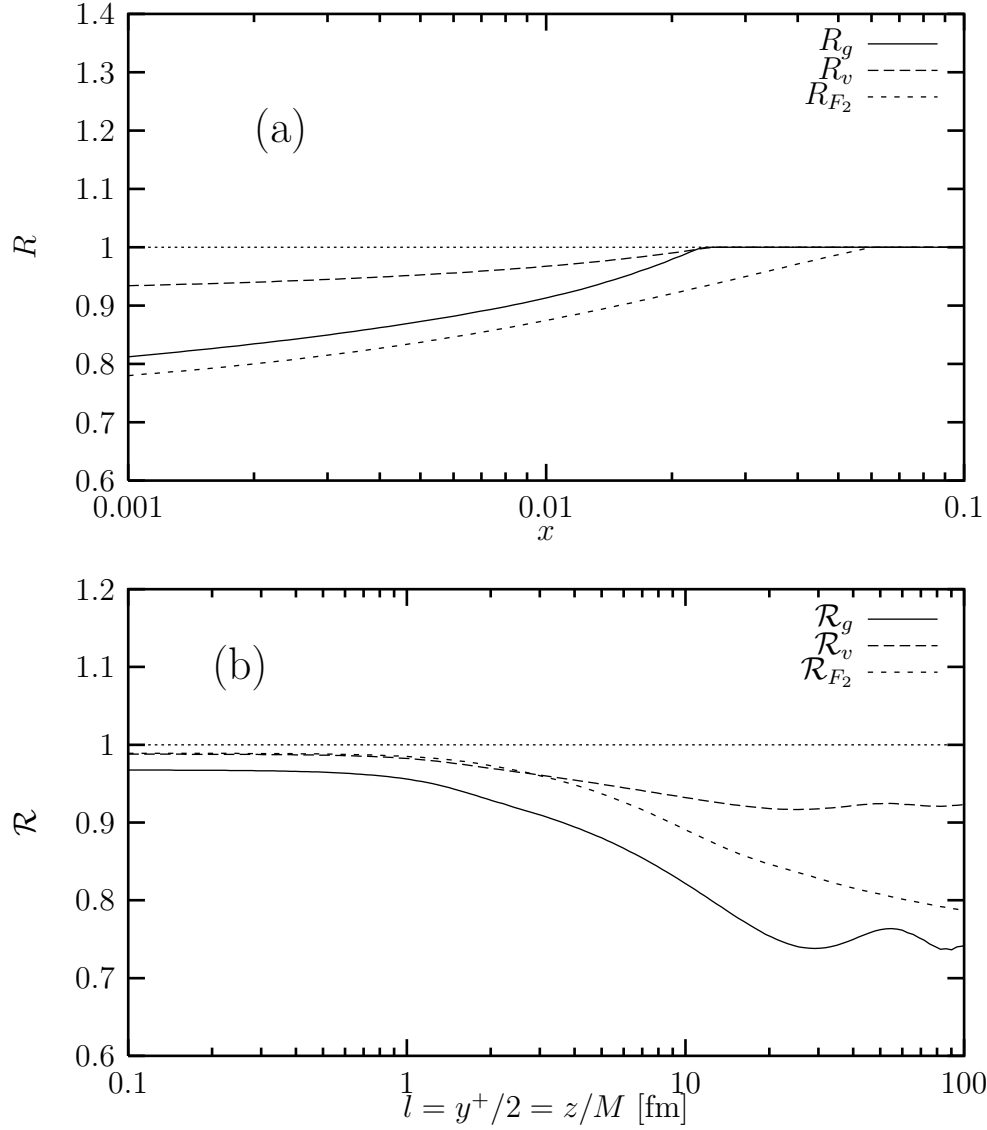

\input{mS.Ca.tex}
\input{cS.Ca.tex}
\caption[]{(a) Shadowing effect in $^{40}$Ca in momentum space
according to Ref.~\protect\cite{Eskola}. (b) The coordinate-space
ratios for gluon distributions, valence-quark distributions, and
the $F_2$ structure function in $^{40}$Ca which are obtained after
setting the momentum-space ratio to one outside the shadowing region.}
\label{fig:shad-only}
\end{figure}

 From Fig.~\ref{fig:coordspace-Ca-Pb} we recall that shadowing 
is clearly the dominant effect in coordinate space. To analyze
this further we first choose $R_v$ and $R_{g}$ to coincide with
the results of \cite{Eskola} in the momentum-space shadowing region
and to be equal to 1 outside this region (see Fig.~\ref{fig:shad-only}a).
The resulting coordinate-space ratios are shown in 
Fig.~\ref{fig:shad-only}b. The strong depletion of nuclear
distributions at large distances $l > 10$ fm is determined almost
completely by the momentum-space distribution functions at $x\lsim 0.1$. 

Next we choose $R_v$ and $R_{g}$ to coincide with \cite{Eskola} in
the momentum-space antishadowing region and to equal $1$ elsewhere. 
By analyzing the corresponding coordinate-space distributions we find
that, against naive expectation, the moderate enhancement localized
in momentum space around $x\simeq 0.15$ corresponds to effects in 
coordinate space which spread over a wide range up to $l \sim 10$ fm.

\begin{figure}
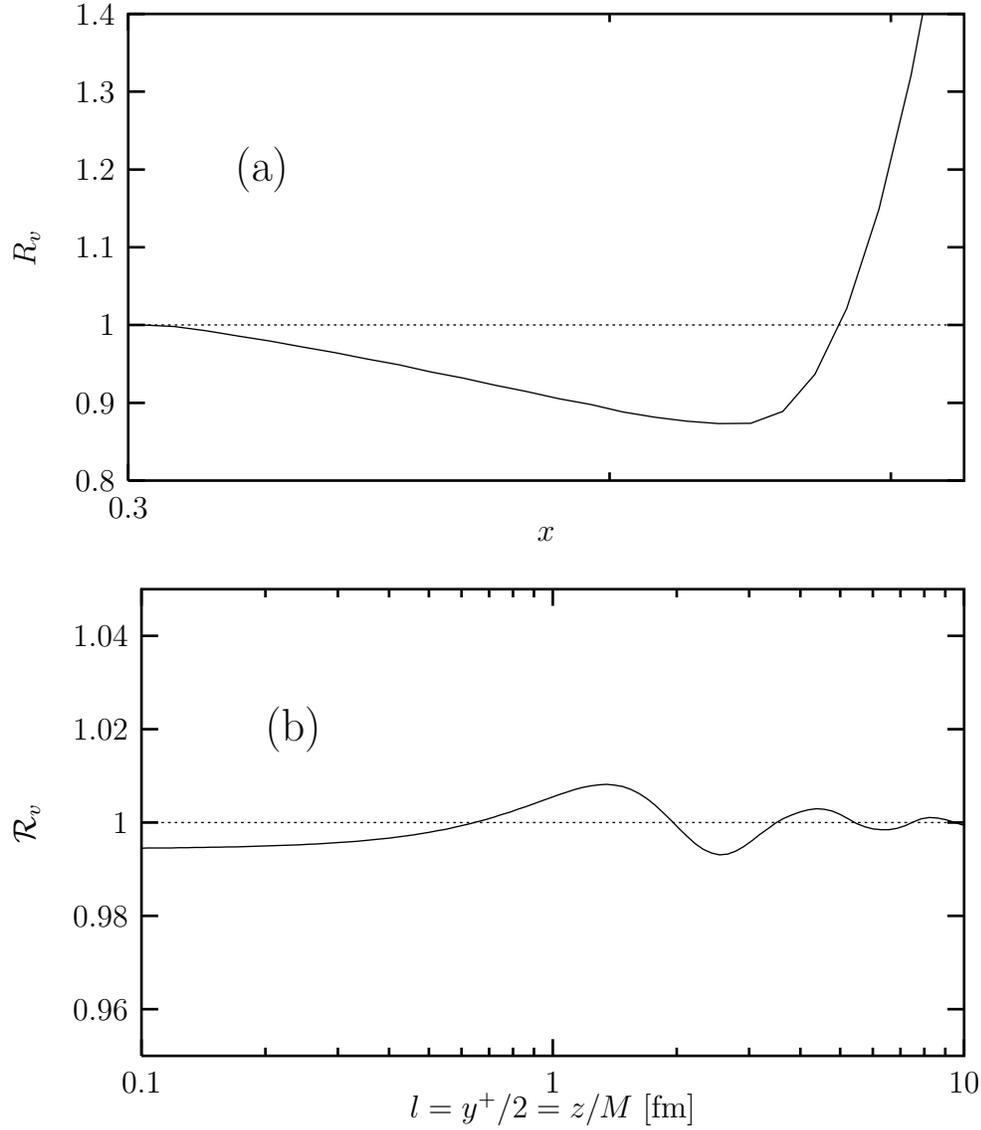

\input{mEF.Ca.tex}
\input{cEF.Ca.tex}
\caption[]{(a) The EMC and Fermi motion effects for valence quarks
in $^{40}$Ca in momentum space according to Ref.~\protect\cite{Eskola}.
(b) The coordinate-space ratio for the valence-quark distribution in
$^{40}$Ca which is obtained after setting the momentum-space ratio
to one below the EMC region.}
\label{fig:EMCandFermi}
\end{figure}

The binding and Fermi motion effects shown in Fig.~\ref{fig:EMCandFermi} 
have an exaggerated appearance close to $x \sim 1$ simply because 
the free nucleon distribution function in the denominator of $R_{v}$ 
decreases rapidly there. These effects turn out to be extremely small
when looked at in coordinate space (see Fig.\ref{fig:EMCandFermi}b). 
Clearly, the combined effect of binding and Fermi motion is marginal 
and leads at most to small fluctuations at the level of $1\%$. 

\section{Summary \label{section:summary}}

In this paper we have investigated nuclear quark and gluon
distributions in coordinate space
\cite{Llewellyn,Hoyer-MV,McLerran}. Coordinate-space parton
distributions are defined as correlation functions involving
two quark or gluon fields separated by a light-cone 
distance $y^+$. To study nuclear effects, the
corresponding longitudinal distance $l = y^+/2$ has to be
compared with typical length scales in nuclei. 

The most significant nuclear effects occur at large longitudinal
distances. A strong depletion of nuclear parton distributions  
(shadowing) is found, starting at $l=2$ fm which corresponds to
the average nucleon-nucleon distance in the bulk of nuclei. In particular
we find that shadowing sets in at approximately the same distance
$l$ for all sorts of partons. This is different in momentum space,
where shadowing starts at different values of Bjorken $x$ for
different distributions.

The magnitude of the shadowing effect increases steadily between
$2$ and $10$ fm, but there is no trace of geometrical boundary
conditions associated with the nuclear radius. Instead, the
shadowing effect continues to increase for distances clearly larger
than the nuclear radius. This feature is also seen in a number of
models for nuclear shadowing (see e.g. \cite{shad-mod} and
references therein). 

At $l \lsim 1$ fm, nuclear modifications of parton distributions
are very small. The intrinsic structure of individual nucleons is
evidently not very much affected by nuclear binding. This
observation becomes even more apparent by looking at the combined
binding and Fermi motion effects. Their influence on nuclear
coordinate-space distributions is at the level of about $1\%$. 

Finally, our analysis does not show evidence for a significant
increase of quark or gluon correlation lengths in bound nucleons.

\end{document}